\documentclass[%
 reprint,
 amsmath,amssymb,
 aps,
]{revtex4-1}

\usepackage{graphicx}
\usepackage{dcolumn}
\usepackage{bm}
\usepackage{color}
\usepackage{soul}
\usepackage[normalem]{ulem}

\begin{document}

\title{Effects of polydispersity on the glass transition dynamics of aqueous suspensions of soft spherical colloidal particles}

\author{Sanjay Kumar Behera, Debasish Saha, Paramesh Gadige and Ranjini Bandyopadhyay}
\affiliation{Soft Condensed Matter Group, Raman Research Institute, C. V. Raman Avenue, Sadashivanagar, Bangalore 560 080, INDIA
}%



\date{\today}

\begin{abstract}
Thermoresponsive poly({\it N}-isopropylacrylamide) (PNIPAM) particles of a nearly constant swelling ratio and with polydispersity indices (PDIs) varying over a wide range ($7.4$\% - $48.9$\%) are synthesized to study the effects of polydispersity on the dynamics of suspensions of soft PNIPAM colloidal particles. The PNIPAM particles are characterized using dynamic light scattering (DLS) and scanning electron microscopy (SEM). The zero shear viscosity ($\eta_{0}$) data of these colloidal suspensions, estimated from rheometric experiments as a function of the effective volume fraction $\phi_{eff}$ of the suspensions, increases with increase in $\phi_{eff}$ and shows a dramatic increase at $\phi_{eff}=\phi_{0}$. The data for $\eta_{0}$ as a function of $\phi_{eff}$ fits well to the Vogel-Fulcher-Tammann (VFT) equation. It is observed that increasing PDIs results in increasingly fragile supercooled liquid-like behavior, with the parameter $\phi_{0}$, extracted from the fits to the VFT equation, shifting towards higher $\phi_{eff}$. The observed increase in fragility is attributed to the prevalence of dynamical heterogeneities (DHs) in these polydisperse suspensions, while the simultaneous shift in $\phi_{0}$ is ascribed to the decoupling of the dynamics of the smallest and largest particles. Finally, it is observed that the intrinsic nonlinearity of these suspensions, estimated at the third harmonic near $\phi_{0}$ in Fourier transform oscillatory rheological experiments, increases with increase in PDIs. Our results are in agreement with theoretical predictions and simulation results for polydisperse hard sphere colloidal glasses and clearly demonstrate that jammed suspensions of polydisperse colloidal particles can be effectively fluidized with increasing PDIs. Suspensions of these particles are therefore excellent candidates for detailed experimental studies of the effects of polydispersity on the dynamics of glass formation.   
\end{abstract}

\pacs{Valid PACS appear here}
\maketitle

 
\section{Introduction}

Thermoresponsive poly({\it N}-isopropylacrylamide) (PNIPAM) hydrogel suspensions undergo a reversible volume phase transition above the lowest critical soluble temperature (LCST) of ${\approx}$ $32$ $^{o}$C in water \cite{Heskins_J_Macromol_Sci_1968, Hirokawa_JCP_1984}. This property of PNIPAM hydrogels provides an opportunity to tune the particle size, volume fraction ($\phi$) and inter-particle interactions in suspension by changing the temperature, thus enabling the study of the complex phase behaviours of soft colloidal suspensions \cite{Wu_Macromolecules_2003, Appel_Softmatter_2016}. Furthermore, suspensions of PNIPAM colloidal particles have been in focus for their applications in controlled and self regulated drug delivery \cite{L. Dong_J_controlled_release_1991}, bioconjugation \cite{Hoffman_clinical_chemistry},  biosensors \cite{Zhang_Sensors_actuators_Chemical}, pollution control \cite{Morris_JCIS_1997}, as viscosity modifiers of liquids and semisolids and in enhanced oil recovery \cite{Sun_softmatter_2011} and chemical separation.

The zero-shear viscosity $\eta_{0}$ of a suspension of hard sphere colloids characterized by a significant polydispersity index (width of the size distribution/mean size $\approx$ 10\%) increases significantly when its volume fraction $\phi$ increases above $\phi\approx0.53$, with the system entering a colloidal glassy state at $\phi_{g}\approx0.58$ \cite{Pusey_PRL_1987, Week_Science_2000}. The dynamics of the colloidal particles freeze at $\phi_{g}$ due to the kinetic constraints imposed on their motion. In order to understand the dramatic slowdown of the dynamics and the formation of disordered glassy phases, the use of colloidal suspensions having a  significant polydispersity index (PDI $\approx10$\%) is essential to prevent the system from entering a crystalline phase \cite{Schaertl_statistical-physics_1994}. The increase in viscosity of fragile glass-forming molecular liquids with decreasing temperature ({\it T}) is well described by the Vogel-Fulcher-Tammann (VFT) law \cite{Angell_JNCS_1991}. For colloidal glasses, the inverse of the volume fraction ($1$/$\phi$) of the colloidal suspension plays the role of temperature ({\it T}) \cite{Brambilla_PRL_2009, Pusey_Nature_1986, Mattsson_Nature_2009}. To use the VFT law for colloidal glasses, therefore, {\it T} is replaced with $1$/$\phi$. The modified VFT equation, $\eta=\eta_{0}\exp\Big(\frac{D\phi}{\phi_{0}-\phi}\Big)$, explains the rise in the viscosity of a colloidal suspension with $\phi$ and its dramatic increase at {$\phi=\phi_{0}$} \cite{Brambilla_PRL_2009, Mattsson_Nature_2009}. Here, $1/D$ is the fragility and accounts for the deviation of the viscosity from an Arrhenius dependence on $\phi$ as the sample approaches the glassy state. In the modified VFT equation, $\eta_{0}=\eta$ ($\phi=0$). A super-Arrhenius dependence of viscosity on the appropriate control parameter (1/{\it T} or $\phi$) is an ubiquitous characteristic of a fragile glass former \cite{Angell_Science_1995}. 

It is known that the increase in  width of the particle size distribution in a mixture of fine powders (fly-ash and silica powder) increases the packing efficiency due to the decrease in the void fraction of the tapped powder beds \cite{Suzuki_2001}. Experiments and a self consistent theory have shown that polydispersity affects the linear viscoelasticity of concentrated polymer solutions and melts \cite{Colby_JCP_1988}. In triblock copolymer Pluronic F127 suspensions, it is shown that the addition of an anionic surfactant, sodium dodecyl sulfate (SDS), breaks the large spherical micelles into smaller, anisotropic polydisperse micelles \cite{Basak_EPJE_2011}. This speeds up the dynamics of the constituents of these mixtures and results in more efficient packing configurations, with the particle volume fraction for random close packing exceeding the value for suspensions of monodisperse hard spheres.

The transition between different arrested states of concentrated binary colloidal mixtures characterized by large size asymmetries has been well-documented in the literature \cite{Yunker_PRL_2010, Hendricks_PRE_2015, Sentjabrskaja_softmatter_2013}. In these experiments, a significant polydispersity value, which is essential to form colloidal glasses, is maintained by mixing the colloidal particles of two different sizes in a certain ratio.

The colloidal glass transition has also been studied for colloidal suspensions comprising particles of significant size polydispersity without mixing batches of monodisperse colloidal particles. The colloidal glass transition of moderately polydisperse hard sphere colloids is reported to smear out with increase in polydispersity \cite{Brambilla_PRL_2009}. The presence of heterogeneous dynamics among the small particles, together with the decoupling of the dynamics of the smallest and largest particles in jammed suspensions of polydisperse hard sphere colloids, shifts $\phi_{g}$ to higher values \cite{Zaccarelli_softmatter_2015}. Recently, Sollich {\it et al.} have reported that the phase boundaries of fluid, fluid-solid and solid phases of hard sphere colloids move to higher $\phi$ with increase in polydispersity \cite{Sollich_softmatter_2011}. Even though there are several simulation results and theoretical models that investigate the effects of polydispersity on the random close packing volume fraction $\phi_{rcp}$ of hard sphere colloids \cite{Schaertl_statistical-physics_1994, Farr_JCP_2009}, these results are difficult to verify experimentally because of the difficulty in systematically controlling the polydispersities of particles synthesized in the laboratory. 

Particle polydispersity is an extremely crucial parameter in determining the fragilities and critical volume fractions of soft glassy systems. Local structural correlations in supercooled liquids, which have been shown in simulations to vary with polydispersity, can profoundly influence the glass transition \cite{Abraham_PRE_2008}. The number of cooperatively coupled molecules ($N_{corr}$), or the size of DHs, can be obtained by analysing the response at the third harmonic to the applied sinusoidal perturbation \cite{Tarzia_JCP_2010}. The sizes of the DHs, or alternatively, the numbers of molecules or particles showing correlated movement ($N_{corr}$), are expected to grow as the glass transition is approached \cite{Thibierge_PRL_2010}. The intrinsic nonlinearity of the polydisperse colloidal system at the third harmonic also grows as the glass transition is approached \cite{Seyboldt_softmatter_2016}. A detailed study of the approach of the suspension to a non-ergodic phase in the presence of varying degrees of polydispersity is therefore important to obtain a clear understanding of the jamming transition. Both the stiffnesses and the polydispersities of soft particles are expected to affect the particle dynamics in suspensions near the colloidal glassy state. Since there is a conspicuous absence of detailed and systematic experimental studies of the dependence of the colloidal glass transition on the size polydispersity of the colloidal particles in suspension, we investigate the glass formation phenomenon in aqueous suspensions of soft PNIPAM particles by changing the particle polydispersity in a controlled manner, while keeping the crosslinker concentration constant to ensure approximately constant particle stiffnesses.   

In the present study, PNIPAM particles characterized by a wide range of PDI values (7.4\%-48.9\%) are successfully synthesized by using one-pot (OP) and semi-batch (SB) methods \cite{Still_JCIS_2013, McPhee_J_Colloid_Interface_Sci_1993}. In OP method, PNIPAM particles of lower polydispersities are synthesized by varying the sodium dodecyl sulfate (SDS) concentration, while in SB method, size polydispersity of the particles is controlled by controlling the feeding rates of the monomer, co-monomer and crosslinker solutions into the reaction vessel. These simple methods are used to control the polydispersity of soft colloidal suspensions and to provide an easy alternative for synthesizing highly polydisperse particles without mixing batches of monodisperse PNIPAM particles. Dynamic light scattering (DLS) and scanning electron microscopy (SEM) are performed to study the thermoresponsivity and the size distributions of these particles. 

The dynamic flow behaviors of the suspensions of soft PNIPAM colloids of varying polydispersities and with increasing particle volume fractions are studied in rheological experiments. Since these particles are soft and deformable, we quantify their packing in aqueous suspensions in terms of an effective volume fraction $\phi_{eff}$. Zero shear viscosity $\eta_{0}$ values, which are extracted from these measurements as a function of $\phi_{eff}$, are observed to increase with increase in $\phi_{eff}$ and shows a dramatic increase at $\phi_{eff}=\phi_{0}$. The $\eta_{0}$ {\it vs}. $\phi_{eff}$ data are analysed using the VFT law to calculate $D$ and $\phi_{0}$ values for suspensions of different $\phi_{eff}$. Furthermore, the intrinsic nonlinearity of these suspensions at the third harmonic is measured from the stress response to oscillatory strains for several colloidal suspensions comprising particles over a range of PDIs. Our results, which can be explained in terms of numerical and analytical studies of polydisperse hard sphere colloidal systems \cite{Schaertl_statistical-physics_1994, Farr_JCP_2009, Yang_PRE_1996, Zaccarelli_softmatter_2015}, demonstrate that the particle packing behaviour, and therefore the glass transition of colloidal suspensions, can be controlled effectively by controlling the polydispersities of the particles in suspension.

\section{Materials and methods}
\subsection{Synthesis of polydisperse PNIPAM particles by the one-pot (OP) method}

PNIPAM particles of lower polydispersities were prepared by the classical OP method. All the chemicals were purchased from Sigma-Aldrich. In a typical experiment, $7.0$ g {\it N}-isopropylacrylamide (NIPAM) ($\geq 99\%$), $0.7$ g {\it N,N'}-methylenbisacrylamide (MBA) ($\geq 99.5\%$) and $0.03$ g SDS were dissolved in $470$ mL of Milli-Q water (Millipore Corp.) in a three-necked round-bottomed (RB) flask fitted with a reflux condenser, a temperature sensor and a $N_{2}$ inlet/outlet. The RB flask was placed on a magnetic stirrer with heating. The solution was stirred at 600 rpm and purged with $N_{2}$ gas for $30$ min to remove oxygen dissolved in water. The temperature of the heat bath was raised to $70$$^{\circ}$C. The free radical polymerization reaction was initiated by the addition of $0.28$ g of potassium persulfate (KPS) ($99.99\%$) dissolved in $30$ mL of Milli-Q water in the preheated bath at $70$$^{\circ}$C. The reaction was allowed to proceed for $4$ hours with a constant stirring speed using a magnetic stirrer. After $4$ hours, the reaction was stopped and the colloidal suspension was cooled down rapidly in an ice bath. The summary of the latex recipes is shown in table T$1$ of Supporting Information. PNIPAM particles of different sizes ($0.16$ $\mu$$m$ - $0.88$ $\mu$$m$) and PDIs ($7.4$\% - $16.2$\%) were synthesized by varying the concentration of SDS in the RB flask. A detailed discussion on how changing the concentration of SDS changes the PDIs of the PNIPAM particles synthesized in this work can be found in page 2 of the Supporting Information.

\begin{table*}
\begin{center}
\begin{tabular}{ |c|c|c|c|c|c| }	
\hline
$Method$ & SDS conc./Flow rate & $<$$d_{25^{\circ}C}$$>$ & $\sigma$($\mu$$m$) & PDI & $\alpha$ \\
\hline	
OP & $0.3$ g/L & $0.16$ $\mu$$m$ & $0.0119$ & 7.4\% & $1.72\pm0.01$\\
\hline	
OP & $0.06$ g/L & $0.40$ $\mu$$m$ & $0.0464$ & 11.6\% & $1.71\pm0.01$\\
\hline	
OP & $0.01$ g/L & $0.47$ $\mu$$m$ & $0.0635$ & 13.5\% & $1.69\pm0.01$\\
\hline
OP & $0.005$ g/L & $0.88$ $\mu$$m$ & $0.1424$ & 16.2\% & $1.68\pm0.01$\\
\hline
SB & $1.5$ mL/min & $1.24$ $\mu$$m$ & $0.1900$ & 15.3\% & $1.81\pm0.02$\\
\hline
SB & $1.0$ mL/min & $1.73$ $\mu$$m$ & $0.5000$ & 28.9\% & $1.87\pm0.01$\\
\hline
SB & $0.7$ mL/min & $2.38$ $\mu$$m$ & $0.8690$ & 36.5\% & $1.84\pm0.01$\\
\hline
SB & $0.5$ mL/min & $2.78$ $\mu$$m$ & $1.3595$ & 48.9\% & $1.83\pm0.02$\\
\hline
\end{tabular}
\caption{Mean hydrodynamic diameters ($<$$d_{H}$$>$), standard deviation of particle sizes ($\sigma$) at $25$$^{\circ}$C, polydispersity indices (PDI) and swelling ratios ($\alpha$) measured from dynamic light scattering experiments at $\theta$=$90$$^{\circ}$ for suspensions of PNIPAM particles synthesized by varying the SDS concentration in the one-pot (OP) method and by varying the flow rate of the reaction ingredients in the semi-batch (SB) method.}
\label{table:Polydispersity-particle_size}
\end{center}
\end{table*}

\subsection{Synthesis of polydisperse PNIPAM particles by the semi-batch (SB) method}

PNIPAM particles of higher polydispersities were prepared in an SB method. $7.0$ g of NIPAM, $0.46$ g of MBA and $34.4$ g of $2$-aminoethylmethacylate hydrochloride (AEMA) were dissolved in $200$ mL of water in a three-necked round bottom (RB) flask fitted with a reflux condenser, a temperature sensor and a $N_{2}$ inlet/outlet. Homogeneously crosslinked PNIPAM particles of different sizes ($1.24$ $\mu$$m$ - $2.78$ $\mu$$m$) and PDIs ($15.3$\% - $48.9$\%) were synthesized by varying the feeding rates of monomer, crosslinker and AEMA solutions. The three-necked RB flask was placed on a magnetic stirrer with heating. The solution was stirred at $600$ rpm for $20$ min and also purged with nitrogen for $20$ min. $180$ mL of this solution was taken out in a syringe and $120$ mL of water was added to the $20$ mL solution remaining in the flask. The temperature of the heat bath was raised to $80^{\circ}$C and the bath was purged with $N_{2}$ gas for $20$ min. The free radical precipitation reaction was initiated by the addition of $41.6$ mg ammonium persulfate (APS) dissolved in $8$ mL of water. Primary nucleation sites started to form after about $5$ min, following which the solution present in the syringe was fed into the reaction vessel at a constant flow rate (between $0.5$ and $1.5$ ml/min) using a syringe pump (Chemyx Inc.). The reaction was stopped $5$ min after all the ingredients were added. The latices were cooled down rapidly in an ice bath. The preparation of the latex recipes is summarized in table T$1$ of Supporting Information. PNIPAM particles of different sizes ($1.24$ $\mu$$m$ - $2.78$ $\mu$$m$) and PDIs ($15.3$\% - $48.9$\%) were synthesized by varying the flow rates of the reaction ingredients into the RB flask. A detailed discussion on the effects of flow rates of the reaction ingredients on the PDIs of the PNIPAM particles synthesized here can be found in page 2 of the Supporting Information.

Latices obtained from SB and OP methods were purified by four successive centrifugations, decantations and redispersions in Milli-Q water to remove SDS, the remaining monomers, oligomers and impurities. Centrifugation was performed at a speed of $11,000$ rpm for $60$ min. Dried polymers were collected by evaporating water from both the supernatant and residue. Fine PNIPAM powders were prepared by crushing and grinding the dried polymers with a mortar and pestle. 

\subsection{Dynamic Light Scattering:} The sizes and size distributions of the PNIPAM suspensions and the thermosensitive responses of these suspensions were determined using dynamic light scattering (DLS) experiments. DLS experiments were performed using a Brookhaven Instruments Corporation (BIC) BI-200SM spectrometer. The details of the setup are given elsewhere \cite{Debasish_YMJ_Ranjini_Soft_Matter_2014}. A temperature controlled water bath (Polyscience Digital) attached to the DLS setup was used to maintain the temperature of the suspension. 
A Brookhaven Instruments BI-9000AT digital autocorrelator was used with the DLS setup to generate the intensity autocorrelation function $g^{(2)}(q,t)$ of the scattered light. Here, $g^{(2)}(q,t) = \frac{<I(q,0)I(q,t)>}{<I(q,0)>^{2}} =  1+ A|g^{(1)}(q,t)|^{2}$ \cite{Bern_Pecora}, where $q$, $I(q,t)$, $g^{(1)}(q,t)$ and $A$ are the scattering wave vector, the intensity at a particular $q$ and a delay time $t$, the normalized electric field autocorrelation function and the coherence factor respectively. $q$ is related to the scattering angle $\theta$ by the equation $q=(4\pi n/\lambda)\sin(\theta/2)$, where $n$ and $\lambda$ are the refractive index of the medium and the wavelength of the laser respectively. For a suspension of monodisperse particles, the normalized electric field autocorrelation function has an exponential decay: $g^{(1)}(q, t) = [\exp(-t/\tau)]$. Here, $\tau$ is the relaxation time of the particle which is directly related to its diffusion coefficient $D_{0}$ by $1/(\tau q^{2})=D_{0}$. The hydrodynamic radius $R_{h}$ of the particle can be calculated using the Stokes-Einstein relation $R_{h}=\frac{k_{B}T}{6\pi\eta D_{0}}$ \cite{Bern_Pecora, Einstein_1905}.
Here $k_{B}$, $T$ and $\eta$ are the Boltzmann constant, the absolute temperature and the viscosity of the solvent respectively.

The decays of the autocorrelation functions estimated for the samples studied here can be fitted to stretched exponential functions, $C(t)=\frac{g^{(2)}(q,t)-1}{A}=\big[\exp(-t/\tau)^{\beta} \big]^{2}$. Here, $A$ is used to normalize the autocorrelation data to $1$ at $t=0$. The stretching exponent $\beta$ is used to calculate the mean and the second moment of the relaxation time spectrum of the particles in suspension using the relations $<$$\tau$$>$$=$$(\frac{\tau}{\beta})\Gamma(1/\beta)$ and $<$$\tau^{2}$$>$$=$ $(\frac{\tau^{2}}{\beta})\Gamma(2/\beta)$ respectively \cite{Lindsey_JCP_1980}, where $\Gamma$ is the Euler Gamma function. The mean hydrodynamic diameter $<$$d_{H}$$>$ and the standard deviation or width of the particle size distribution $\sigma$ were obtained by using $<$$d_{H}$$>$$=\frac{k_{B}T<\tau>q^{2}}{3\pi\eta}$ and $\sigma=\frac{k_{B}T\sqrt{<\tau^{2}>-<\tau>^{2}}q^{2}}{3\pi\eta}$ respectively \cite{Segre_PRL_1996}. 

For the DLS measurements reported here, PNIPAM suspensions of a very dilute concentration ($c=0.01$ wt\%) were prepared by adding dried PNIPAM powder in Milli-Q water. $5.0$ mL of each suspension was filled in a glass cuvette and the cuvette was placed in the sample holder of the DLS setup attached to a temperature controller. The swelling and shrinkage behaviors of PNIPAM particles were characterized at various temperatures at $3$$^{\circ}$C intervals in the range $20$$^{\circ}$C - $50$$^{\circ}$C during cooling and heating. The swelling ratio $\alpha$ of the PNIPAM particles, calculated by measuring the mean hydrodynamic diameter $<$$d_{H}$$>$ of the particle in the completely swollen state ($<$$d_{20^{\circ}C}$$>$) and in the completely shrunken state ($<$$d_{45^{\circ}C}$$>$), is defined as $\alpha$$=$$<$$d_{20^{\circ}C}$$>$$/$$<$$d_{45^{\circ}C}$$>$ (table $1$). 

The normalized electric field autocorrelation function $g^{(1)}(q, t)$ is related to the line width distribution $G(\Gamma)$ by the relation $g^{(1)}(q, t)=$$<$$E(t, q)E^{*}(0, q)$$>$$=\int_{0}^{\infty} G(\Gamma)\exp(-\Gamma t)d\Gamma$ \cite{Koppel_JCP_1972, Xia_langmuir_2004}. Here, $\Gamma$ is the decay rate which is related to $D_{0}$ by the relation $\Gamma=D_{0}q^{2}$. The particle size distribution of each PNIPAM particle suspension is obtained from the inverse Laplace transform of $g^{(1)}(q, t)$ \cite{Koppel_JCP_1972, Xia_langmuir_2004}. The values of mean relaxation time $<$$\tau$$>$ and mean hydrodynamic diameter $<$$d_{H}$$>$, obtained by fitting the intensity autocorrelation data of a PNIPAM suspension synthesized in the OP method at an SDS concentration of 0.06 g/liter (figure S$1$(a) of Supporting Information), and in the SB method at a flow rate of $0.7$ mL/min of the flow rates of the reaction ingredients(figure S$1$(b) of Supporting Information), are seen to decrease with the increase in temperature (table T$2$ of Supporting Information).
The size distribution of PNIPAM particles is quantified by a dimensionless polydispersity index (PDI) which is defined as the ratio of the width of the particle size ($\sigma$) and the mean hydrodynamic diameter ($<$$d_{H}$$>$) of the particle.

\subsection{Scanning Electron Microscopy:} PNIPAM particles were directly visualized using a scanning electron microscope (SEM, GEMINI column, ZEISS, Germany). An ITO substrate was cleaned with acetone and a PNIPAM suspension of concentration $0.01$ wt\% was added on the ITO coated side of the substrate using a pipette. The sample placed on the ITO substrate was dried overnight at $25$$^{\circ}$C and was then loaded on the SEM stage for imaging. The electron beam interacts with the atoms of the sample and the backscattered secondary electrons are used to produce the surface images of the samples. There were some particle doublets and triplets that were visible in the images. These were not considered in the size analysis. The average particle sizes and size distributions were estimated from approximately $1,000-1500$ particles.

\subsection{Rheology:}
Rheological measurements were performed with a stress controlled Anton Paar MCR $501$ rheometer. For these measurements, PNIPAM suspensions of very high concentrations were prepared by adding dried PNIPAM powder in Milli-Q water. The suspension was then stirred for $24$ hours and sonicated for $45$ minutes. PNIPAM suspensions of different concentrations were prepared by diluting suspensions of the highest PNIPAM particle concentration with Milli-Q water. A concentric cylinder geometry (CC-17) of measuring bob radius $r_{i}=8.331$ mm, measuring cup radius $r_{e}=9.041$ mm, cone angle $\alpha=120^{\circ}$ and gap length $L=24.990$ mm was used for dilute samples. For concentrated samples, a cone-plate geometry (CP-25) of cone radius $r_{c}=12.491$ mm, cone angle $\alpha=0.979^{\circ}$ and measuring gap $d=0.048$ mm was used. The inner surfaces of concentric cylinder and cone-plate geometries are made of smooth stainless steel. A solvent trap oil was used to minimize solvent evaporation. The temperature of the samples were maintained constant at $25$$^{\circ}$C using a Peltier unit and a water circulation system for counter cooling. For each experiment, a sample volume of $4.7$ mL was loaded in the concentric cylinder geometry, while a sample volume of $0.07$ mL was loaded in the cone-plate geometry. Measurements with both roughened (sandblasted) and smooth geometries produced comparable results and verify the absence of wall slip in the smooth geometries used in this work (figure S$2$ of Supporting Information).
		
Steady state flow experiments of PNIPAM suspensions were performed at $25$$^{\circ}$C. The viscosity $\eta$ of each suspension was measured by varying the shear rate $\dot{\gamma}$ between $0.001$ and $4000$ $s^{-1}$. The zero-shear viscosity $\eta_{0}$ of each suspension was calculated by fitting the viscosity $\eta$ versus shear rate $\dot\gamma$ curve (figure S$3$ of Supporting Information) with the Cross model \cite{Cross_JCS_1965}:
\begin{equation}
\frac{\eta-\eta_{\infty}}{\eta_{0}-\eta_{\infty}}=\frac{1}{1+(k\dot{\gamma})^{m}}
\end{equation}
where $\eta_{0}$ and $\eta_{\infty}$ correspond to the viscosity plateaus at very low and very high shear rates respectively, $k$ is a time constant related to the relaxation time of the polymer in solution and $m$ is a dimensionless exponent. The relative viscosity $\eta_{rel}$ of a PNIPAM suspension is defined as the ratio of its zero-shear viscosity $\eta_{0}$ and the viscosity of water $\eta_{s}$: $\eta_{rel}=\eta_{0}/\eta_{s}$.

The PNIPAM particles are soft and start deforming in the presence of neighbouring particles at volume fractions above the random close packing volume fraction of undeformed monodisperse spheres ($\phi_{RCP}=0.64$) \cite{Pusey_Nature_1986, Mattsson_Nature_2009}. Since the volume fraction {$\phi$} does not account for the particle deformations of these soft particles, we consider a modified parameter called the effective volume fraction $\phi_{eff}$. $\phi_{eff}$ of a PNIPAM suspension was obtained from the relation $\phi_{eff}=nV_{d}$, where $n$ is the number of particles per unit volume and $V_{d}=\pi$$<$$d_{H}$$>$$^{3}/6$ is the volume of an undeformed particle of mean hydrodynamic diameter $<$$d_{H}$$>$ in a very dilute suspension \cite{Mattsson_Nature_2009}. The above relation can be written in terms of the polymer concentration of the microgel packing $c$ as $\phi_{eff}=c/c_{p}$ \cite{Lorenzo_macromolecules_2013}, where $c_{p}=m_{p}/V_{d}$ is the polymer concentration inside each particle in the swollen state and $m_{p}$ is the molecular weight of each particle. The Batchelor's equation relates $\eta_{rel}$ of the suspension with its $\phi_{eff}$ as follows: \cite{Batchelor_JFM_1997,Brady_J.Rheology_1995}:
\begin{equation}
\eta_{rel}=1+2.5\phi_{eff}+5.9\phi_{eff}^{2}
\end{equation}
The polymer mass concentration of the suspension can be converted into the effective volume fraction by substituting $\phi_{eff}=c/c_{p}$ in equation 2: 
\begin{equation}
\eta_{rel}=1+2.5(c/c_{p})+5.9(c/c_{p})^{2}
\end{equation}
The Batchelor's law is valid for only dilute and relatively monodisperse hard sphere suspensions \cite{Batchelor_JFM_1997}. 

In our experiments, only dilute concentrations ($\phi_{eff}$ $<$ 0.2) of polydisperse PNIPAM suspensions are used to rule out inter-particle interactions \cite{Poon_colloidal_suspension}. Our assumption is validated by the excellent fits of the data to equation (3) (figure S4 of Supporting Information). The values of $c_{p}$, calculated by fitting equation (3) to the experimental data displayed in figure S$4$ of the Supporting Information, are tabulated in table T$3$ for PNIPAM particles of different mean hydrodynamic diameters $<$$d_{H}$$>$ and polydispersity indices (PDIs). The fitted values of the polymer concentration inside each particle, $c_{p}$, is independent of the polymer mass concentration of the suspension $c$ and depends only on the particle preparation protocol. $\phi_{eff}=c/c_{p}$ is therefore used to determine the value of the effective volume fraction $\phi_{eff}$ at higher concentrations for the same batch of particles.

Medium amplitude oscillatory shear (MAOS) tests were performed for suspensions of PNIPAM particles of lower PDIs synthesized by the OP method using the CP-$25$ geometry. PNIPAM suspensions of different volume fractions near critical volume fractions $\phi_{c}$ were prepared using several different batches of particles. The Fourier coefficients of the elastic and viscous moduli at the third harmonic,  $G^{\prime}_{3}$($\omega$,$\gamma_{0}$) and $G^{\prime\prime}_{3}$($\omega$,$\gamma_{0}$), and at the fundamental harmonic, $G^{\prime}_{1}$($\omega$,$\gamma_{0}$) and $G^{\prime\prime}_{1}$($\omega$,$\gamma_{0}$), were measured by varying the strain amplitude $\gamma_{0}$ between 0.1\% to 50\% at a fixed angular frequency $\omega$. The intrinsic moduli [$G^{\prime}_{3}$($\omega$)] and [$G^{\prime\prime}_{3}$($\omega$)] at a very low strain amplitude were determined by fitting the strain amplitude dependence of the Fourier coefficients of the elastic and viscous moduli at the third harmonic, $G^{\prime}_{3}$($\omega$,$\gamma_{0}$) and $G^{\prime\prime}_{3}$($\omega$,$\gamma_{0}$), with quadratic functions, $G^{\prime}_{3}=[G^{\prime}_{3}(\omega)]\gamma_{0}^{2}$ and $G^{\prime\prime}_{3}=[G^{\prime\prime}_{3}(\omega)]\gamma_{0}^{2}$ (figure S$5$ of Supporting Information) \cite{Seyboldt_softmatter_2016}. The frequency dependent equilibrium linear elastic and viscous moduli $G^{\prime}_{eq}(\omega)$ and $G^{\prime\prime}_{eq}(\omega)$ were determined by fitting the strain amplitude dependence of the fundamental Fourier coefficients of the elastic and viscous moduli, $G^{\prime}_{1}$($\omega$,$\gamma_{0}$) and $G^{\prime\prime}_{1}$($\omega$,$\gamma_{0}$), to the equations $G^{\prime}_{1}=G^{\prime}_{eq}(\omega)+[G^{\prime}_{1}(\omega)]\gamma_{0}^{2}$ and $G^{\prime\prime}_{1}=G^{\prime\prime}_{eq}(\omega)+[G^{\prime\prime}_{1}(\omega)]\gamma_{0}^{2}$ (figure S$6$ of Supporting Information). The intrinsic nonlinearity $Q_{0}(\omega)$ was obtained from the relation $Q_{0}(\omega)=\frac{|[G_{3}(\omega)]|}{|G_{eq}(\omega)|}$ \cite{Seyboldt_softmatter_2016}. Here, $|[G_{3}(\omega)]|=[[G^{\prime}_{3}(\omega)]^{2}+[G^{\prime\prime}_{3}(\omega)]^{2}]^{1/2}$ and $|G_{eq}(\omega)|=[[G^{\prime}_{eq}(\omega)]^{2}+[G^{\prime\prime}_{eq}(\omega)]^{2}]^{1/2}$ are the magnitudes of the intrinsic moduli at the third harmonic and the frequency dependent equilibrium linear complex modulus respectively. 

We observe some difference in the quality of the $G^{\prime}_{3}$ and $G^{\prime\prime}_{3}$ data acquired from suspensions of PNIPAM particles synthesized by OP and SB methods at very low $\gamma_{0}$ (figure S$7$ of Supporting Information). It is difficult to determine the intrinsic parameters $[G^{\prime}_{3}(\omega)]$ and $[G^{\prime\prime}_{3}(\omega)]$ for the highly polydisperse PNIPAM suspensions synthesized by the SB method due to the presence of large fluctuations in the $G^{\prime}_{3}$ and $G^{\prime\prime}_{3}$ values measured at very low $\gamma_{0}$ (figure S$7$(b) of Supporting Information). These observed fluctuations arise from the inhomogeneities present in these highly polydisperse suspensions. Since the data from the suspensions of PNIPAM particles synthesized by the OP method are less noisy (figure S$7$(a) of Supporting Information), we therefore use their aqueous suspensions to study the nonlinear Fourier transform rheology at the first higher harmonic.  
\begin{figure}[!b]
\includegraphics[width=3.4in]{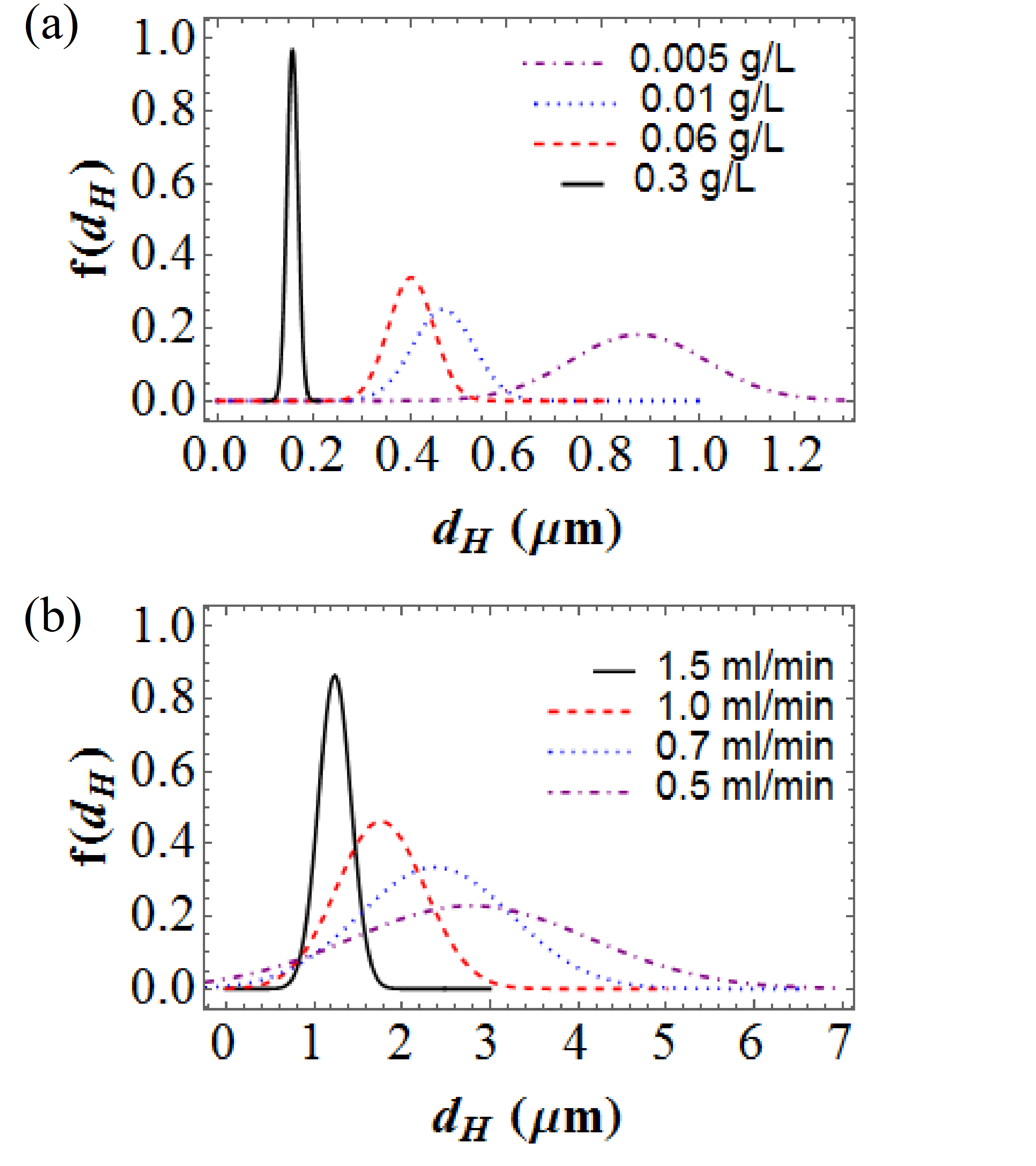}
\caption{(a) The hydrodynamic particle size distribution of PNIPAM particles measured from DLS experiments for particles synthesized by the OP method at different SDS concentrations (g/liter) at $25$$^{\circ}$C and at a scattering angle $\theta$=$90$$^{\circ}$. (b) The hydrodynamic particle size distribution of PNIPAM particles measured from DLS experiments for particles synthesized by the SB method at different flow rates (ml/min) of the reaction ingredients at $25$$^{\circ}$C and at a scattering angle $\theta$=$90$$^{\circ}$.}  
\label{Particle_size_distribution-DLS}
\end{figure}   

\section{RESULTS AND DISCUSSIONS}
\subsection{PNIPAM particles characteristics:}
The plots of hydrodynamic particle size distributions, estimated from DLS at a scattering angle $\theta$=$90$$^{\circ}$, for PNIPAM particles synthesized by the one-pot (OP) and the semi-batch (SB) methods are shown in figure $1$. The sizes ($<$$d_{H}$$>$), swelling ratios ($\alpha$) and polydispersity indices (PDIs), estimated for all the synthesized PNIPAM particles in aqueous suspensions in the DLS experiments, are listed in table 1. It is seen from figure 1 and table $1$ that the mean and the width of the size distribution of the PNIPAM particles increase with the decrease in concentration of SDS. Using the OP method, we have synthesized particles of PDIs varying from 7.4\% to 16.2\%. Particles of higher polydispersity (PDIs between 15.3\% and 48.9\%) are synthesized in the SB method by varying the flow rate of the reaction ingredients . It is seen from figure 1 and table 1 that the decrease in the flow rate of the reaction ingredients in the SB method increases the mean and the width of the size distribution of the PNIPAM particles. The increase in mean particle sizes and PDIs with decreasing SDS concentration in the OP method and with decreasing flow rates of the reaction ingredients in the SB method have been verified by making additional measurements at lower scattering angles (figure S$8$ of Supporting Information). The range of PDI values of the soft PNIPAM particles synthesized here varies between 7.4\% and 48.9\%. This wide range  allows us to systematically investigate the effects of PDI values on the approach of concentrated PNIPAM suspensions towards the colloidal glass transition.
\begin{figure}[!b]
\includegraphics[width=3.4in]{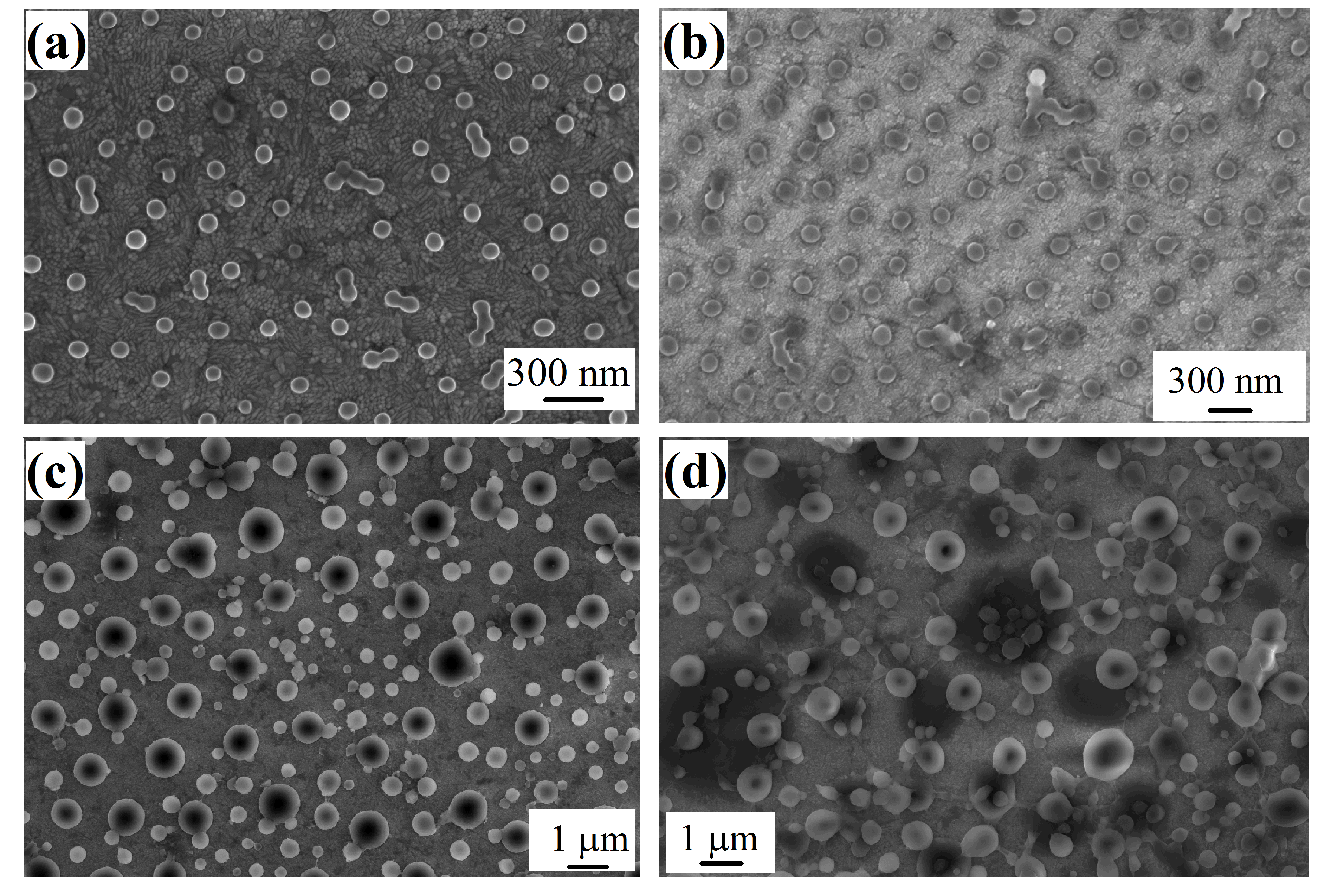}
\caption{(a-b) Representative SEM micrographs of dry PNIPAM particles synthesized by the OP method (a) particle PDI 11.9\% at SDS concentration of $0.06$ g/liter and (b) particle PDI 16.2\% at SDS concentration $0.01$ g/liter. (c-d) Representative SEM micrographs of dry PNIPAM particles, synthesized by the SB method (c) particle PDI 46.9\% at a flow rate of $0.7$ mL/min and (d) particle PDI 53.8\% at a flow rate of $0.5$ mL/min.}
\label{SEM_images-EP-SB}
\end{figure}

Direct visualization of the PNIPAM particles synthesized by the OP and the SB methods is achieved using SEM imaging. Representative SEM images are displayed in figure~\ref{SEM_images-EP-SB}. The average sizes and particle size distributions estimated from SEM images are shown in table T1 and figure S$9$ of the Supporting Information. These results are in excellent agreement with the DLS results (table~\ref{table:Polydispersity-particle_size} and figure$1$). However, the sizes of the particles measured from SEM images are found to be less than those obtained by DLS at $45$$^{\circ}$C by a factor of almost $1.2-2.0$  (table T$1$ of Supporting Information). This is explained by considering that above the LCST ($45$$^{\circ}$C), the particles in suspension shrink by expelling water from their interior, but are still not collapsed completely. The DLS measurements therefore reveal larger values of $<$$d_{H}$$>$ when compared to the SEM experiments in which the particles are in the dried state.

The particle stiffness is characterized by measuring the temperature dependence of the sizes of the PNIPAM particles using DLS. The corresponding thermoresponsive curves of the PNIPAM particles in suspension are shown in figure 3. It is seen that the swelling ratio $\alpha$ is approximately insensitive to the method of synthesis (figure $3$). This is explained by considering that a fixed concentration of crosslinker MBA is used in both the OP and SB methods (table T$1$ of Supporting Information). It can therefore be concluded that the swelling behaviour, and therefore, the CLD inside each particle is approximately insensitive to changes of PDI.
\begin{figure}[!t]
\includegraphics[width=3.7in]{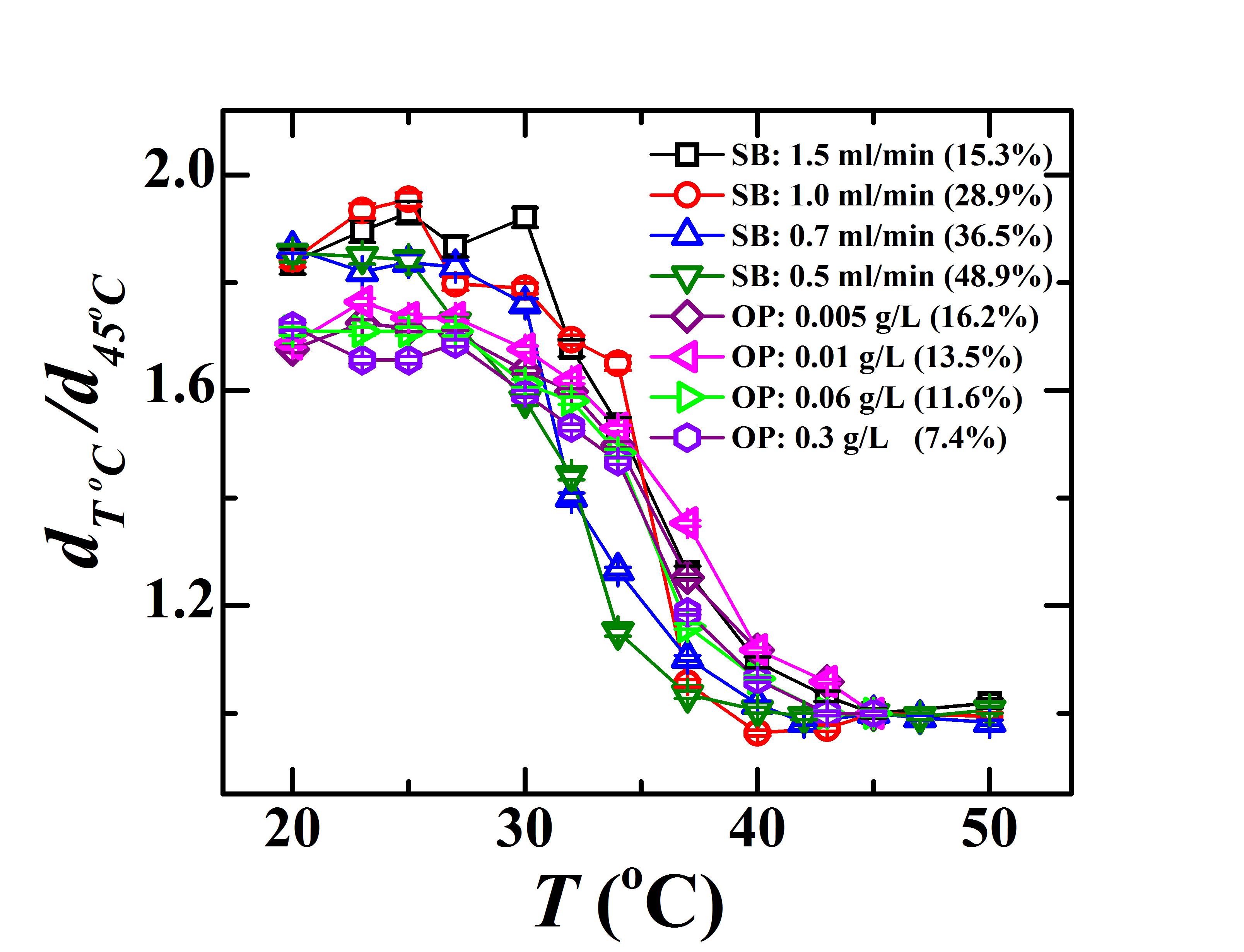}
\caption{Temperature dependent swelling ratio $d_{T^{\circ}C}/d_{45^{\circ}C}$ as a function of temperature $T^{\circ}C$ for suspensions of PNIPAM particles of different PDIs synthesized by the SB method at different flow rates (ml/min) of the reaction ingredients, and by the OP method, synthesized using different concentrations (g/L) of SDS, estimated from DLS experiments at a scattering angle $\theta$=$90$$^{\circ}$.}  
\label{swell_particle_diameter-EP}
\end{figure}
\begin{figure}[!t]
\includegraphics[width=3.4in]{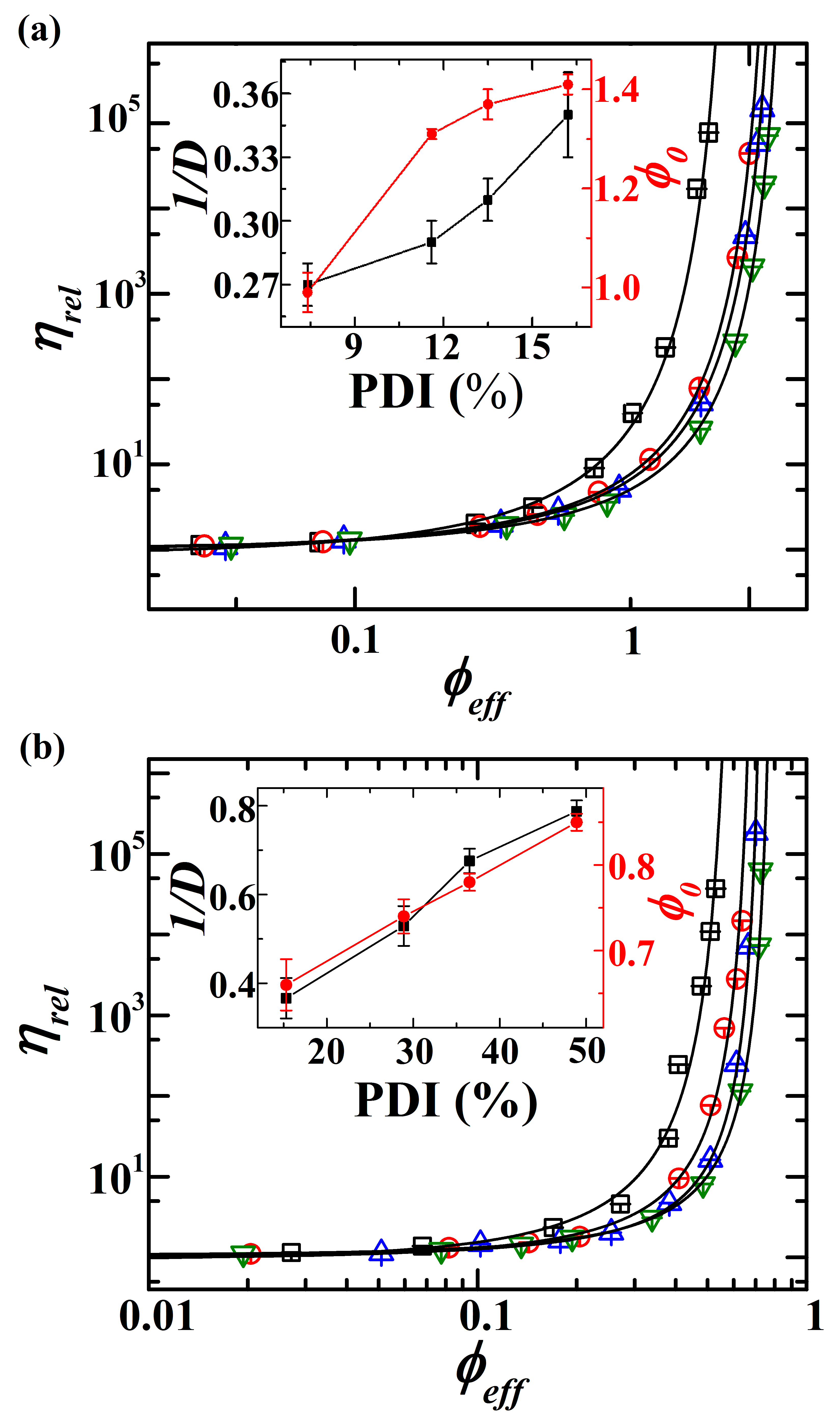}
\caption{(a) Relative viscosity $\eta_{rel}$ {\it vs.} effective volume fraction $\phi_{eff}$ of suspensions of PNIPAM particles synthesized by the OP method at different concentrations of SDS and characterized by different PDIs: 7.4 $\%$ ($\Box$), 11.6 $\%$ ($\circ$), 13.5 $\%$ ($\triangle$), 16.2 $\%$ ($\nabla$) at $25^{\circ}$C. (b) $\eta_{rel}$ {\it vs.} $\phi_{eff}$ of suspensions of PNIPAM particles synthesized by the SB method at different flow rates of reaction ingredients and characterized by different PDIs: 15.3 $\%$ ($\Box$), 28.9 $\%$ ($\circ$), 36.5 $\%$ ($\triangle$), 48.9 $\%$ ($\nabla$) at $25^{\circ}$C. The solid lines are the VFT fits (equation $4$) to the data. Insets show the fitted values of fragility $1/D$ ($\blacksquare$) and $\phi_{0}$ ({\color{red}$\bullet$}) {\it vs.} PDI of the PNIPAM particles in suspension.}  
\label{relative_viscosity-vol_fraction-EP}
\end{figure}

\subsection{Glass formation dynamics of polydisperse PNIPAM suspensions:}
The dynamics of glass formation as a function of PDIs in suspensions of soft PNIPAM particles having PDIs between $7.4\%$ and $48.9\%$ is investigated using rheology at $25^{\circ}$C. The colloidal glass transition is expected to occur at a volume fraction $\phi_{g}$ at which the viscosity shows an apparent divergence. It is well known that for colloidal glasses, the volume fraction $\phi$ plays a role analogous to the inverse of the temperature $T$, with the dynamics slowing down with increasing $\phi$ due to the increase in kinetic constraints \cite{Hunter_RPP_2012, Megen_PRL_1993}. Since $\phi_{eff}$ is a correct measure of the concentration of soft PNIPAM particles in aqueous suspensions, we estimate the relative zero shear viscosity $\eta_{rel}$ as a function of $\phi_{eff}$. $\eta_{rel}$ {\it vs.} $\phi_{eff}$ plots are depicted in figure $4$. $\eta_{rel}$ values are observed to increase by four decades over the range of $\phi_{eff}$ explored here. 

The values of $\eta_{rel}$ of suspensions of soft PNIPAM particles in the arrested state are higher than the $\eta_{rel}$ values reported for hard sphere colloidal particle suspensions \cite{Meeker_PRE_1997, Phan_PRE_1996}. The observed higher values of $\eta_{rel}$ in these suspensions of polydisperse, deformable PNIPAM particles approaching $\phi_{g}$ are due to the increase in the number of contacts and the number of neighbours surrounding a given particle, which result in the enhancement of the particle packing efficiency. The values of the fragility, $1/D$, and the effective volume fraction $\phi_{0}$, at which $\eta_{rel}$ increases dramatically, are obtained by fitting the $\eta_{rel}$ {\it vs.} $\phi_{eff}$ data (figure $4$) to the VFT equation \cite{Mattsson_Nature_2009, Brambilla_PRL_2009}:

\begin{equation}
\eta_{rel}=\exp\Big(\frac{D\phi_{eff}}{\phi_{0}-\phi_{eff}}\Big)
\end{equation}
Here, $D$ is the fragility parameter. The fragility ($1/D$) in glass-forming liquids is defined as the deviation of the viscosity from Arrhenius growth upon cooling rapidly (for supercooled liquids) or with increasing $\phi$ (for colloidal suspensions) \cite{Angell_Science_1995, Tanaka_PRL_2003}.

The values of $1/D$ and $\phi_{0}$, obtained from fits to the VFT law for suspensions of PNIPAM particles with varying PDIs are reported in table ~\ref{table:Polydispersity-critical_volume_fraction}. It is seen that $1/D$ and $\phi_{0}$ increase with increase in PDIs. 
\begin{table}
\begin{center}
\begin{tabular}{ |c|c|c|c|c|c|c| }
\hline
Method & PDI & $1/D$ & $\phi_{0}$ \\
\hline
OP & $7.4\%$  & $0.27\pm 0.01$ & $0.99\pm 0.04$ \\
\hline
OP & $11.6\%$  & $0.29\pm 0.01$ & $1.31\pm 0.01$ \\
\hline
OP & $13.5\%$  & $0.31\pm 0.01$ & $1.37\pm 0.03$ \\
\hline
OP & $16.2\%$  & $0.35\pm 0.02$ & $1.41\pm 0.02$ \\
\hline
SB & $15.3\%$  & $0.37\pm 0.05$ & $0.66\pm 0.03$ \\
\hline
SB & $28.9\%$  & $0.53\pm 0.04$ & $0.74\pm 0.02$ \\
\hline
SB & $36.5\% $ & $0.68\pm 0.03$ & $0.78\pm 0.01$ \\
\hline
SB & $48.9\% $ & $0.79\pm 0.02$ & $0.85\pm 0.01$ \\
\hline
\end{tabular}
\caption{Fragility $1/D$ and $\phi_{0}$ values obtained by fitting the VFT equation to the relative viscosity $\eta_{rel}$ {\it vs}. effective volume fraction $\phi_{eff}$ data (figure $4$) for suspensions of PNIPAM particles of different PDIs.} \label{table:Polydispersity-critical_volume_fraction}
\end{center}
\end{table}
The observed increase of $\phi_{0}$ with increasing PDIs for the PNIPAM particle suspensions studied here is in agreement with theoretical predictions and simulation results for close packed hard sphere colloids \cite{Schaertl_statistical-physics_1994, Farr_JCP_2009, Yang_PRE_1996} and can be explained by considering that the packing of smaller spherical particles in between bigger spherical particles becomes increasingly efficient as the sample PDI increases. It has been shown that suspensions of polydisperse hard spheres can stay fluidized at volume fractions higher than the random close packing volume fraction of monodisperse hard spheres \cite{Zaccarelli_softmatter_2015, Brambilla_PRL_2009}. This is due to the high degree of heterogeneous dynamics prevalent among the small particles whose mobilities are higher than those of the bigger particles in the polydisperse mixture \cite{Zaccarelli_softmatter_2015}. Heterogeneous dynamics, and the resultant decoupling of the dynamics of the small and big particles in the polydisperse suspensions, result in the big particles forming an arrested state, with the smaller particles of higher diffusivities smearing out the glass transition to a certain degree \cite{Zaccarelli_softmatter_2015}. The increased prevalence of heterogeneous dynamics and the increased packing efficiency of PNIPAM microparticles with increase in the polydispersity index can therefore explain the data in figure~\ref{relative_viscosity-vol_fraction-EP}.

It is seen from table 2 that the $\phi_{0}$ values for suspensions of particles of lower polydispersities, synthesized by the OP method, are higher than those of the suspensions of the particles synthesized by the SB method (table ~\ref{table:Polydispersity-critical_volume_fraction}). This can be attributed to the rigid cores and loose shells of the particles produced in the OP method \cite{Saunders_Langmuir_2004}. The heterogeneous crosslinking density (CLD) of the PNIPAM particles synthesized by the OP method produces highly crosslinked rigid cores, and less crosslinked hairy shells \cite{Pelton_Langmuir_1989}. At sufficiently high volume fractions, the core-shell particles deform up to the core because of the presence of the hairy shells. This shifts $\phi_{0}$ to higher values. The PNIPAM particles synthesized by the SB method, in contrast, have a uniform CLD within each particle \cite{Still_JCIS_2013}. This is confirmed by direct visualization (figure S$10$ of the Supporting Information). The larger polymer densities at the core of the particles obtained in the OP method create a larger refractive index mismatch between the particles and the background solvent (water). This leads to stronger light scattering and more suspension turbidity in suspensions of these particles when compared to the particles synthesized by the SB method. At fixed volume fraction above $\phi_{RCP}$, therefore, the particles synthesized by the SB method deform less when compared to the particles synthesized by the OP method.
\begin{figure*}[!t]
\includegraphics[width=7in]{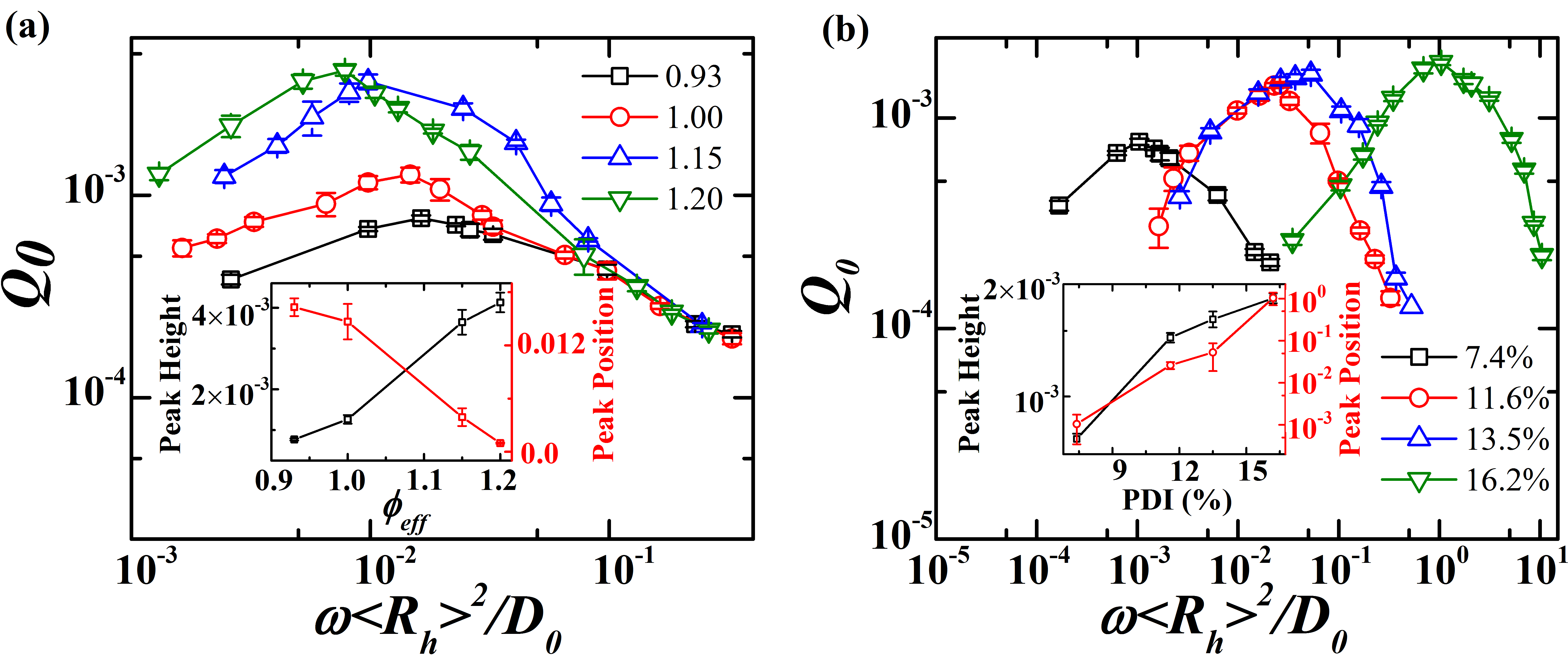}
\caption{(a) Intrinsic nonlinearity $Q_{0}$ at the third harmonic {\it vs.} $\omega<R_{h}>^{2}/D_{0}$ for suspensions of PNIPAM particles synthesized by the OP method for different $\phi_{eff}$ near $\phi_{0}$ at a PDI = 11.6\% at $25^{\circ}$C. (b) Intrinsic nonlinearity $Q_{0}(\omega)$ at the third harmonic {\it vs.} $\omega<R_{h}>^{2}/D_{0}$ for suspensions of PNIPAM particles synthesized by OP method and characterized by different PDIs at a $\phi_{eff}=1.0$ at $25^{\circ}$C. Inset shows the peak height ($\Box$) and peak position ({\color{red}$\circ$}) of $Q_{0}(\omega<R_{h}>^{2}/D_{0})$ curve {\it vs.} PDI of the PNIPAM particles in suspension.}  
\label{Intrinsic_nonlinearity-frequency}
\end{figure*}

The fragility of glass forming liquids is ascribed to the presence of slow and fast moving groups of molecules or particles called dynamical heterogeneities (DHs). The observed high fragility values of these glassy suspensions of polydisperse particles can be understood by considering an increase in the concentration of the DHs \cite{He_PRE_1999, Kiriushcheva_PRE_2001}. Indeed, the fragility values increase by almost a factor of three over the range of PDIs explored in this work (table II and solid squares in the insets of figure $4$(a) and $4$(b)). The enhancement of heterogeneous dynamics with increasing PDIs is further supported by the third harmonic nonlinear response data discussed in the following section.

\subsection{Study of the intrinsic mechanical nonlinear susceptibility of polydisperse PNIPAM suspensions:}
The nonlinear susceptibilities of glass forming liquids have been studied recently to evaluate its dependence on the length scales of the dynamical heterogeneities \cite{Brambilla_PRL_2009, Thibierge_PRL_2010, Ladieu_PRB_2012, Paramesh_Sci-Rep_2017}. Moreover, the nonlinear rheological response of visco-elastic materials have been studied to understand material rigidity and its microscopic origin \cite{Hyun_macromolecules_2009}. A recent report on soft PNIPAM core-shell particle suspensions showed the divergence of the mechanical nonlinear susceptibility at the third harmonic while approaching the glass transition \cite{Seyboldt_softmatter_2016}. In order to understand the heterogeneous behaviour and fragility of polydisperse PNIPAM suspensions, we record the mechanical nonlinear susceptibility arising at the third harmonic in the stress response of concentrated polydisperse PNIPAM suspensions using Fourier transform oscillatory rheology. To estimate the nonlinearity of these suspensions, we calculate the intrinsic nonlinearity $Q_{0}$, which is the ratio of the mechanical susceptibilities at the third harmonic with respect to the first harmonic at a very low strain amplitude $\gamma_{0}$. $Q_{0}$ is estimated as a function of angular frequency $\omega$ of the applied strain for PNIPAM suspensions synthesized by the OP method at several different $\phi_{eff}$ near $\phi_{0}$ at a fixed PDI value and for several different PDIs at a fixed $\phi_{eff}$.

In figures $5(a)$ and figure $5(b)$, $Q_{0}$ is plotted as a function of $\omega$, normalized by the intrinsic relaxation frequency ($D_{0}/<R_{h}>^{2}$) of a single particle, obtained from dilute suspensions for PNIPAM suspensions at several different $\phi_{eff}$ at a fixed PDI and for several different PDIs at a fixed $\phi_{eff}$, respectively. $Q_{0}$ exhibits a peak with increase in $\omega<R_{h}>^{2}/D_{0}$ in both plots. It is seen from figure 5(a) that the peak height increases by half a decade and the peak position shifts to lower frequencies in PNIPAM suspensions when $\phi_{eff}$ increases from 0.93 to 1.20 at a fixed particle PDI. The values of the peak heights and the peak positions at various $\phi_{eff}$ are plotted in the inset of figure 5(a). When the PDIs of the particles in these suspensions are increased from $7.4\%$ to $16.2\%$ at a fixed $\phi_{eff}$, the peak height is observed to increase by a half a decade, while the peak positions shift to higher frequencies (figure $5(b)$). The values of the peak positions and the peak heights {\it vs}. PDI are plotted in the inset of figure 5(b). The plots of $Q_{0}$, estimated as a function of $\omega<R_{h}>^{2}/D_{0}$ for PNIPAM suspensions of different $\phi_{eff}$ near $\phi_{0}$ at a PDI = 11.6\%, are seen to collapse at high frequencies (figure $5(a)$). This observation indicates the presence of a single fast relaxation time in PNIPAM suspensions of different $\phi_{eff}$ values near $\phi_{0}$ when the particle PDI is kept constant. However, in figure $5(b)$, the plots corresponding to different PDIs at fixed $\phi_{eff}$ are observed to have different shapes and therefore cannot be collapsed onto a master curve. This suggests the presence of different relaxation rates in the PNIPAM suspensions having different PDIs at a fixed $\phi_{eff}$.  

The observed increase in the peak heights with increase in $\phi_{eff}$ at a fixed PDI (inset of figure 5(a)) and with increase in PDI at a fixed $\phi_{eff}$ (inset of figure 5(b)) confirm an increase in the intrinsic nonlinearities of the suspensions when the relevant control parameter ($\phi_{eff}$ or PDI) is changed. This can be interpreted in terms of increasingly heterogeneous dynamics which is also apparent from the fragility measurements reported in figure 4. In general, for suspensions characterized by higher particle PDIs, the dynamics of small particles decouple from the dynamics of big particles, which leads to a broadening of the relaxation spectra of the particles. This is expected to increase the heterogeneity in the dynamics of suspensions of higher particle PDIs.

Furthermore, the shift of the peak position to a higher frequency with increase in particle PDI indicates a decrease in the structural relaxation times (inset of figure 5(b)) of the particles in aqueous suspension. This can be explained by considering that the smaller particles in the highly polydisperse suspensions still remain free to diffuse even at very high volume fractions. This results in the emergence of an arrested state at larger particle concentrations when compared to suspensions having particles of smaller PDIs. From the nonlinear rheology results shown in figures $5(a)$ and $5(b)$, we can therefore separate the effects of changes in $\phi_{eff}$ and PDI on the glass transition of PNIPAM suspensions. Increase in $\phi_{eff}$ at a fixed PDI accelerates the jamming of PNIPAM particles in suspension, while increase in PDI at a fixed $\phi_{eff}$ fluidizes the jammed state of PNIPAM suspensions. Qualitatively, the increase in $Q_{0}$ with increase in particle PDI indicates an increase in the concentrations and length scales of dynamical heterogeneities and supports the viscosity data shown in figure 4.

\section{Conclusions}
We have synthesized thermoresponsive PNIPAM particles of a wide range of polydispersity indices (PDIs) using the one-pot and semi-batch synthesis methods. The thermoresponsive behaviour and the size distributions of these particles are characterized by DLS and SEM. A rheological study of the approach of these suspensions towards the glass transition shows that the soft and polydisperse PNIPAM particles form highly fragile glasses at volume fractions well above the random close packing volume fraction $\phi_{rcp}$ of monodisperse hard sphere colloidal suspensions. The nonlinear stress response at the third harmonic to applied oscillatory strains is measured to quantify the intrinsic nonlinearity of these samples. The intrinsic nonlinearity of these suspensions is monitored by changing $\phi_{eff}$ at a fixed PDI and by changing particle PDIs at a fixed $\phi_{eff}$. The nonlinearity of the dynamics of these suspensions is observed to increase with increase in both $\phi_{eff}$ and PDI. The increase in fragility and $\phi_{0}$ of these concentrated suspensions with increase in particle PDI is supported by our nonlinear rheology results. For a jammed packing of polydisperse particles, the local packing fraction, the number of contacts and the number of neighbours surrounding a given particle are the parameters that decide the packing efficiency at the single particle level. All these parameters are expected to increase with increase in $\phi_{eff}$, PDI and the particle softness, thereby enhancing the packing efficiency of the soft and deformable PNIPAM particles. Suspensions of PNIPAM particles are therefore excellent candidates for studying the dynamics of highly polydisperse particles in soft glassy suspensions.    

\section{Acknowledgements}
We thank A. Dhason and K. M. Yatheendran for their help with SEM and cryo-SEM imaging.

\end{document}